# Thermal stability and long term hydrogen/deuterium release from soft to hard amorphous carbon layers analyzed using *in-situ* Raman spectroscopy. Comparison with Tore Supra deposits.

C. Pardanaud[1*], C. Martin[1], G. Giacometti[1], N. Mellet[1], B. Pégourié[2], P. Roubin[1]

[1] Aix-Marseille Université-CNRS, PIIM, 13397 Marseille cedex 20, France.

[2] CEA, IRFM, 13108 Saint-Paul-lez-Durance, France

\* Corresponding Author: cedric.pardanaud@univ-amu.fr

**Abstract**

The thermal stability of 200 nm thick plasma enhanced chemical vapor deposited a-C:H and a-C:D layers ranging from soft to hard layers has been studied and compared to that of deposits collected on the Tore Supra tokamak plasma facing components by means of in-situ Raman spectroscopy. Linear ramp heating and long term isotherms (from several minutes to 21 days) have been performed and correlations between spectrometric parameters have been found. The information obtained on the $sp^2$ clustering has been investigated by comparing the G band shift and the 514 nm photon absorption evolution due to the thermal treatment of the layer. The effects of isotopic substitution have also been investigated.

**Key words:** Raman spectroscopy, Long term hydrogen release, Amorphous carbon, Thermal stability

# 1. Introduction

Understanding and controlling hydrogen isotope retention in materials is of importance for nuclear fusion applications. In fusion devices (tokamak), a magnetically confined D+T plasma is heated to produce energy. In these devices, few eV to hundreds of eV ions bombard the inner walls, leading to plasma/wall interactions in which implantation, heating, diffusion, outgassing, deposition, etc., can occur depending on the local physical conditions. Due to H-rich deposits consecutive to plasma/wall interactions, hydrogen retention (essentially tritium) is a key issue for the next step fusion device ITER because the total inventory in the machine is limited to 700 g for safety reasons.

For decades, carbon was widely used for the tokamak walls and will still be in use in the future machines JT60-SA (tokamak) and W7-X (stellarator). Although - for safety issue - carbon, which forms a-C:H layers, has been left away as promising material for ITER, the experience gained in the understanding of the processes at play in plasma/wall interactions is still useful. For example, detailed analyses of D retention in the Tore Supra tokamak carbon walls have been performed, leading to a consistent understanding of the D balance and migration during the whole Tore Supra operation, quantifying in details the amount of D in a-C:D deposits [1-4]. This study has also revealed the importance of the long term D-release occurring between the phases of plasma operation, which needs to be taken into account to reconcile post mortem and particle balance analyses [1, 2]. Note that a similar long term D-release has recently been observed due to the beryllium walls in the JET tokamak, in the ITER like wall configuration [5].

Recent studies dealing with the thermal stability and long term hydrogen release of a-C:H layers [6, 7] have allowed to investigate how useful information, such as structure evolution and hydrogen release, can be retrieved from Raman spectra. It is known that Raman spectroscopy is a powerful tool to characterize C-based materials that contain $sp^2$ carbon atoms ($C(sp^2)$) such as graphite, graphene, nanotubes, soots, etc. The 1000 - 1800 cm$^{-1}$ spectral window, which gives information on the carbon organization (aromatization), is dominated by two bands: the G and D bands [8]. These bands contain information on the disorder such as the size of aromatic domains, the density of defects… For amorphous carbons (a-C), these bands can give an indication on the $C(sp^2)/C(sp^3)$ ratio, and the H content for hydrogenated carbons (a-C:H) [8-11]. Spectroscopic parameters generally used to probe the bonding structure are: the width and the position of the G band, denoted $\Gamma_G$ and $\sigma_G$ respectively, the height ratio of the D and G bands, $H_D/H_G$, and the $m/H_G$ parameter, m being the slope of the photoluminescence background.

We have recently shown that, for heat-treated a-C:H layers, we have measured a linear relation between $\log_{10}(m/H_G)$ and the H content (H/H+C = 25 + 9 $\log_{10}(m/H_G)$) in the range H/H+C = 15 to 30 %, meaning that this spectroscopic parameter is sensitive to the H bonded to $C(sp^3)$ [7]. This is similar to the relation that was previously obtained for as-deposited samples for which the H content has been varied from 20 to 47 at. % [12, 13]. We have emphasized that $m/H_G$ is sensitive not only to the H content, but also to defect passivation [14, 15]. In addition, we have obtained a linear relation between $H_D/H_G$ and the H content (H/H+C = 0.54 - 0.53 $H_D/H_G$) in the range H/H+C = 2 to 30 at.%, meaning that this spectroscopic parameter is sensitive to H bonded to $C(sp^2)$ and to $C(sp^3)$. Conversely, we have shown that the parameters $\Gamma_G$ and $\sigma_G$ are mainly sensitive to the structural changes such as the re-organization leading to a larger aromatization, and the loss of $C(sp^3)$ induced by heating. In [6], we have evidenced the long term kinetics of the isothermal H release from a-C:H layers with different initial H/H+C ratios and various thicknesses, at temperatures of 450 °C, 500 and 600 °C. We have shown how the kinetic evolution (fast regime followed by a slower one) provides valuable information on the H release processes and emphasizes the role of the initial structure and H-content on these processes: the higher the initial H content, the faster the kinetics, in agreement with what is known on the instability of this type of hydrogenated layers.

In this study, the aim is to complete the previous results we obtained on the thermal stability and long term hydrogen/deuterium release of controlled a-C:H and a-C:D thin deposited layers and to compare them with deuterated deposits extracted from the Tore Supra tokamak plasma facing components. We explore the differences and similarities found by using both linear temperature ramps and isotherms. We validate our methodology as a useful non-contact methodology for long term hydrogen isotope release study for fusion applications and believe it will be used in hydrogenated beryllium samples in the future.

## 2. Experimental

Plasma-deposited hydrogenated ($CH_4$ precursor), deuterated ($CD_4$ precursor) and H-free amorphous carbon layers were synthesized and deposited on silicon substrates at IPP Garching, Germany. These layers will be referred to as reference samples in the following. For the as-deposited samples (AD), H contents H/H+C are~ 0.29, 0.32 and 0.37 (DC self-bias of −300, −200 and −100 V, respectively), for thicknesses e ~ 0.24, 0.25 and 0.19 $\mu$m, and an imaginary part of the refractive index k (633 nm)~ 0.097, 0.072 and 0.032 ± 0.005 (see ref [18] for details on the deposition and characterization methods). D-content D/D+C are ~ 0.31, 0.33 and 0.38 (DC self-bias of −300, −200 and −100 V, respectively) for thicknesses e ~ 0.21, 0.19 and 0.21 $\mu$m, and an imaginary part of the refractive index k(633 nm)~ 0.050, 0.035 and 0.027 ± 0.010. These values are summarized in table 1. These layers can be roughly classified as hard or soft layers, according to their H content and optical properties, as speculated in [18], with typical values of hardness ranging from ~10 to 20 GPa for the former and less than 10GPa for the latter, according to [19]. Then a-C:H layers with H/H+C ~ 0.29 and 0.32 are hard layers while that with H/H+C ~0.37 is intermediate between a hard and a soft layer [18, 20]. The a-C:Dlayer with D/D+C ~ 0.31 is a hard layer whereas that with D/D+C ~ 0.33 is intermediate between a hard and a soft layer. The layer with D/D+C ~ 0.38 is a soft layer. Deuterated amorphous samples from the Tore Supra tokamak [21] were collected on the deposition zone of the Toroidal Pumped Limiter (the main plasma facing component of the device) after a period of five years of operation (39 h of cumulated plasma time). Their thicknesses were not measured precisely on the samples analyzed by means of *in-situ* Raman spectroscopy, but they are in the range of tens to hundreds of microns, depending on where they were collected in the machine, according to [22]. Despite their difference in thickness, Tore Supra deposits are porous, allowing desorption mechanisms similar to those at work in the reference samples to occur [23, 24]

Both types of samples were heat treated under 1.5 bar of argon (purity: 99.9995%, from Air liquid company) in a LINKAM TS1500 cell. Linear temperature ramps (T=$\alpha$t+$T_0$, $\alpha$=3 °C.min$^{-1}$, $T_0$ being the room temperature) together with isothermal heatings (ranging from 300 to 600 °C and with a duration from several minutes to up to 21 days) were used. Raman spectra were recorded in situ through a window, except when specifically mentioned (data displayed in figure 4 and 5-c). Raman spectra were obtained using a Horiba-Jobin-Yvon HR LabRAM apparatus (laser wavelength: $\lambda_L$ = 514.5 nm, 50X objective, numerical aperture of 0.5 leading to a laser focus diameter of 2.5 µm, resolution ≈ 1 cm$^{-1}$). The laser power was 1 mW. The analysed Raman parameters were the G band wavenumber, $\sigma_G$, the G band full-width at half-maximum, $\Gamma_G$, the relative height of the G and D bands, $H_D/H_G$, and the $m/H_G$ parameter. For reference samples, we also analysed the $H_{Si}/H_G$ ratio, where $H_{Si}$ is the height of the underlying silicon wafer whose band lies at 520 cm$^{-1}$. A linear background was calculated between 800 and 2000 cm$^{-1}$ (slope m) and subtracted allowing the heights to be measured on the raw data. This last choice was done because a debate still exists in the literature (see the introduction in [25]) concerning how to fit amorphous carbon Raman spectra: by using 4 to 5 Gaussians (with 1 G band and other components labelled D1 to D4) or by including an asymmetric profile for the G band with less D bands. Recent advances obtained for more ordered carbons have shown that a Fano profile for the G band plus two D bands can be justified with good physical reasons [25]. However, as we do not address this question here, we chose the simplest way to analyse the data and we used only raw spectral parameters. $H_D$ was therefore measured at its apparent maximum, except when the D band maximum was not enough well defined (i.e. for AD sample). In this latter case, $H_D$ was taken at 1370 cm$^{-1}$.

## 3. Results and discussion

3.1 Linear temperature ramp study

Figure 1 displays Raman spectra of AD and heat treated a-C:Hsamples(H/H+C ~ 0.29, fig 1-a) and a-C:D (D/D+C ~ 0.31, 0.33 and 0.38, fig 1-b, 1-c and 1-d, respectively) after the linear backgrounds have been subtracted, and intensities normalized to the G band height. Note that for the AD, 200 and 400°C heated samples with D/D+C ~ 0.38, the background is more structured and cannot be approximated by a line between 800 and 2000 cm$^{-1}$ in the 400°C case. We then diminished the range of the spectral window to 1200-1700 cm$^{-1}$, in which we assume that the background can be approximated by a lines. This figure also displays the Raman spectra corresponding to these samples heated from room temperature to 200, 400, 600 and 700°C under argon atmosphere with a heating rate $\alpha$=3°C.min$^{-1}$. All these Raman spectra exhibit two bands which are broad and which overlap. These two bands are the well-known G and D bands, even visible for the AD spectra (in figure 1, only the fit corresponding to the G band is displayed, to show by comparison with the experimental data that the D band is present in the four AD samples). The main trend for these four samples is that the D band intensity increases when the temperature increases, that the G band blue shifts, and that both bands narrow so that they are less overlapping for high temperatures. All these evolutions reveal the material organization under heating, as explained in more details in [6, 7, 9, 26, 27]. A closer look reveals that for T ≥ 400°C, the sample with the lowest H or D content displays the spectrum where the two G and D bands are the broadest and the most overlapping, and where the D band is the less intense. For example, at 400°C, $H_D/H_G$ starts from 0.46 for H/H+C~0.29 and increases up to 0.76 for D/D+C~0.38. Note that the signal to noise ratio can be affected by three parameters: the sample modification (refractive index modification, see figure 3), the layer delamination at high temperature and the laser focalisation variation induced by the thermal gradient existing close to the window's cell.

Figure 2 displays the thermal evolution of $\sigma_G$, $\Gamma_G$, $m/H_G$ and $H_D/H_G$ for 8 samples (the 7 reference samples and one sample from the Tore Supra tokamak) heated with a linear ramp $\alpha$ = 3°C.min$^{-1}$ from room temperature to 800°C. As photoluminescence is more efficient for the D/D+C~0.38 sample than for the others, it is that for which the Raman signature is the less intense. We therefore chose a higher acquisition time by spectrum, which explains why there are less spectra during the ramp when compared to the other samples. Figure 2-a shows that AD frequencies are grouped around three values: $\sigma_G \sim 1545$ cm$^{-1}$ for H free a-C, ~1525 cm$^{-1}$ for a-C:H samples and ~1500 cm$^{-1}$ for a-C:D samples. Therefore, the redshift observed for the AD reference layers (25 cm$^{-1}$ between hydrogenated and deuterated layers) is of the same order of magnitude than that observed for the benzene and the deuterated benzene carbon stretching modes (respectively at 1599 cm$^{-1}$ and 1557 cm$^{-1}$ [28]), suggesting that the observed redshift is mainly due to an isotopic effect. When the temperature increases, $\sigma_G$ of all the samples evolves with a "s" shape. This evolution is interpreted as a diminution of the C(sp$^3$) content and/or as an increase of the size of the aromatic domains [7, 27] (see below for more details). At a specific temperature depending on the sample, $\sigma_G$ reaches a plateau, at ~1570 cm$^{-1}$, which is independent on the sample, revealing either that the size of the aromatic domains has reached a value high enough so that their stretching modes are no longer sensitive to the H or D bonds remaining in the material, or that the degree of order in the material does no longer evolve. The AD soft layer, with D/D+C ~ 0.38, starts to evolve at a lower temperature, reaching the plateau for T=400°C, whereas the AD hardest layer, with H/H+C ~ 0.29 initially, reaches the plateau at 600°C. This observation is consistent with what is known concerning the stability of hard/soft hydrogenated layers [29-33]. The thermal evolution of the hydrogen free a-C has the lowest amplitude, starting from $\sigma_G \sim 1545$ cm$^{-1}$ at room temperature to 1570 cm$^{-1}$ at 800°C. The thermal evolution of the Tore Supra sample is close to that of the H-free a-C, revealing that the deuterium atoms bonded in the material do not modify the sp$^2$ carbon structure. Figure 2-b displays the thermal evolution of $\Gamma_G$ for the same samples. Contrary to the $\sigma_G$ values corresponding to the reference AD samples, the $\Gamma_G$ values are not sprayed around three values, but roughly concentrate around one value close to ~200 cm$^{-1}$, the AD D/D+C ~ 0.38 being an exception as $\Gamma_G$ lies at~220 cm$^{-1}$. When the temperature increases, $\Gamma_G$ of all the samples evolves with a "s" shape, the higher the initial H or D content, the lower the temperature at which $\Gamma_G$ starts to decrease, as for $\sigma_G$. For T>600°C, $\Gamma_G$ for all the samples has reached an own plateau, which is in the range 77-112 cm$^{-1}$. This may be due to some orientation disorder remaining different in each sample. For the Tore Supra deposits, the $\Gamma_G$ value recorded at room temperature has a lower value than the others, which suggests that there is less disorder in the deposits. $\Gamma_G$ starts to decrease close to 300°C.

Figure 2-c shows that for the AD samples, the values are grouped around three values: m/H$_G$~ 2 for the H/H+C~ 0.29 and 0.32 samples, m/H$_G$~ 5 for H/H+C ~ 0.37, D/D+C ~ 0.31 and 0.33, and m/H$_G$~ 20 for D/D+C ~ 0.38. This is in agreement with the hard, intermediate between hard and soft, and soft layers attribution we did looking on optical properties (see experimental section). Between room temperature and 500 °C, there is a gap between the value of m/H$_G$ for the D/D+C~0.33 and the H/H+C~0.32 samples although their H or D contents are very close. As the DC self-bias is the same for these two samples, the kinetic energy of the impinging ions is the same. Then, we explain this isotopic effect in terms of the number of defects created during the deposition process. This number is higher for D than for H due to different momentum transfer at the time of surface impact. For all samples, when the temperature increases, m/H$_G$ increases up to a factor of roughly two before it starts to decrease. The increase of m/H$_G$ is due to an increase of photoluminescence (increasing m) due to aromatization and/or defect passivation occurring at the first stages of heating. Therefore, m/H$_G$ cannot be simply related to the H content in this temperature range, as shown in [7] and references therein. The temperature where

m/$H_G$ decreases, meaning that the H content starts to decrease due to H-C(sp$^3$) breaking bond, is found at ~500°C for hard layers. This is consistent with what is found using infrared spectroscopy [34]. For the intermediate between hard and soft layers, this temperature is found at ~ 420°C whereas it is found at ~380°C for the soft layer. On the one hand, this difference in the thermal stability at low temperature suggests that it is easier to break the H-C(sp$^3$) bonds in soft layers than in hard ones. On the other hand, for temperatures higher than 540°C, the similar evolution of m/$H_G$ for all layers indicates that the remaining H-C(sp$^3$) bonds are identical in all the samples. From this observation, we infer that roughly two kinds of H-C(sp$^3$) bonds may exist in soft layers whereas in hard layer, only one kind exists. This is consistent with the fact that, when changing the DC self-bias from low to high negative values, the number of sp$^3$-CH$_3$ groups decreases while the number of sp$^3$-CH$_2$, sp$^3$-CH, and sp$^2$-CH$_x$ groups increases, according to the infrared analysis made in [35]. For temperatures lower than 680°C, the slope m of a-C is close to zero. For higher temperatures, the black body radiation of the heater becomes significant, changing the background (m/$H_G$ increases from 0.1 at 680°C to 7 at 800°C) and preventing from measuring the intrinsic m/$H_G$ value of hydrogenated and deuterated samples. Note that the Tore Supra sample's thermal evolution of this spectroscopic parameter, contrary to that of the other samples, does not show a maximum but decreases like the other samples although starting at a lower temperature (~250°C). This last point, plus the fact that $\sigma_G$ evolves like that of to the H-free sample, means that D is trapped as C(sp$^3$)-D$_x$ far from the C(sp$^2$) aromatic domains in the Tore Supra samples.

Figure 2-d shows that $H_D/H_G$ for the H-free sample starts at 0.76 and rises up to 0.9 at 800°C, while $H_D/H_G$ for the other samples starts at~0.4 and reaches ~0.9 at 800°C,

suggesting that the thermal evolution of this parameter is mainly driven by the presence of hydrogen or deuterium bonded to carbons. This is in agreement with what we have shown previously [7]. For D/D+C~0.38, $H_D/H_G$ starts to increase at around 150°C whereas it starts at 250-300°C for D/D+C~0.33, 0.31 and H/H+C~0.37 and at 400°C for H/H+C~0.32 and 0.29. For T higher than 680°C, the $H_D/H_G$ thermal evolution becomes independent on the sample. Then the H-C($sp^2$) content remaining in each sample for T> 680°C is independent on the initial hydrogen content. This temperature is also the temperature where we cannot yet measure $m/H_G$, meaning that, at this temperature, the H-C($sp^3$) content is low [6]. For the Tore Supra sample, $H_D/H_G$ starts at a value (~0.5) higher than that of all the other hydrogenated or deuterated samples. At 800°C it reaches ~1, its thermal behaviour being down shifted by~200°C when compared to that of the hard layer samples.

Figure 3-a displays the thermal evolution of $H_{Si}/H_G$, where $H_{Si}$ is the height of the band lying at 520 $cm^{-1}$ (not shown here), which is characteristic of the silicon wafer on which the a-C:H and a-C:D layers have been deposited. The thermal evolution of this spectroscopic parameter corresponding to the hydrogenated and deuterated layers follow these trends: $H_{Si}/H_G$ starts from different values depending on the initial H or D content (it ranges from $H_{Si}/H_G$~0.7 for the ADH/H+C~0.29 sample to $H_{Si}/H_G$~5 for the ADD/D+C~ 0.38 sample), and decreases down to $H_{Si}/H_G$~0.03 for all the reference samples, this value being the detection limit in our experimental conditions. In our conditions, since the thicknesses of all these layers are similar (~200 nm) and since the kind of silicon wafer is the same for all the samples studied, the $H_{Si}/H_G$ spectroscopic parameter corresponding to the AD samples can be used to rapidly estimate their H or D content. The reason for this is consistent with what is known from the literature: it is shown in [18] that the samples which contain the more H

are those with the lowest light absorption coefficient, and are the less thermally stable [32]. However, there is a factor of ~2 between the $H_{Si}/H_G$ value corresponding to the AD D/D+C~0.33 sample and that of the H/H+C~0.32 sample, whereas their initial H or D contents are close. This difference is due to a different number of defects created by mass selective momentum transfer after the ions impinge the surface, as explained above. Figure 3-b displays the imaginary part of the refractive index, k, of the reference samples between room temperature and 600°C, derived from the value of $H_{Si}/H_G$, knowing the k and thickness values of the AD samples and applying the method detailed in [36]. Small grey squares are estimations of the error bars. They have been plotted only for the H/H+C ~ 0.29 data in order to simplify the figure. Error bars for the other samples are estimated to be of the same order. Between room temperature and 590°C, k increases by one order of magnitude for the softersample (D/D+C~0.38), whereas it is increased by a factor of three for the harder sample (H/H+C~0.29). At ~530°C all the samples display a value close to k~0.2, which increases up to k~0.29 in the same way for all the samples up to 590°C. At that temperature, the samples become optically thick and no more evolution of k can be determined in our experimental conditions. In the range 530-590°C all the reference samples continue to absorb light more efficiently when the temperature increases, whereas we have shown above that in that thermal range, $\sigma_G$ has reached a plateau (figure 2-a). The comparison of the thermal behavior of these two parameters allows us to conclude that when the aromatic domains are too large, $\sigma_G$ is no more sensitive to the size increase. This is consistent with the results of [34], where it is found that the studied layers absorb more photons when the clustering/organization increases. According to [20, 34], the thickness may increase due to heat treatment, but this increase is lower than 6% for a temperature increasing from room temperature to 600°C. We then did not took this effect into account in the calculation since k varies by one order of magnitude in that thermal range. One has to note that since the value of $H_{Si}/H_G$ decreases when the temperature increases, the values of k deduced at the highest temperatures become less accurate.

3-2 Isothermal study

Figure 4 displays the Raman data of the Tore Supra samples. Figure 4-a is a mapping of $H_D/H_G$ for an AD sample. The D content is given as an indication using the formula deduced in [7]. One can see that the deposit is inhomogeneous, but the inhomogeneities are few microns large (the lateral resolution of the mapping is ~500 nm), $H_D/H_G$ varying between 0.5 and 0.7, corresponding to a D content variation between 17 and 28 % on this mapping. Figure 4-b and-c display the $\sigma_G$ and $H_D/H_G$ histograms built from the mapping shown in figure 4-a, and those built from the mappings recorded after heating at 120°C and 250°C s. For the AD sample, the average value of $\sigma_G$ is 1537 cm$^{-1}$ whereas it is shifted to higher wavenumbers by 1 cm$^{-1}$ only, at 1538 cm$^{-1}$, for the sample heated at 120°C, and by 21 cm$^{-1}$, at 1558 cm$^{-1}$ for that heated at 250°C. The corresponding average values of $H_D/H_G$ are respectively 0.52, 0.53 and 0.59. The shape is Gaussian for the AD and 120°C-heated samples and starts to become asymmetric for the sample heated at 250°C, the width being twice in this case. Our interpretation is that there are structural and chemical inhomogeneities that evolve with different characteristic times, leading to several components in the histograms. Structural modification and deuterium release are then evidenced at 120°C and higher temperatures, which is in agreement with what was found in[2].

Figure 5 displays the $H_D/H_G$ time evolution under isothermal heating, measured *in-situ* for the reference samples and measured at room temperature for the heated Tore Supra sample (figure 5-a: 400 to 600°C for the H/H+C ~ 0.29 sample, figure 5-b: 350-600°C for the D/D+C ~ 0.31 sample, and figure 5-c: 300-600°C for the Tore Supra sample). Curves of all the reference samples start at $H_D/H_G$ ~ 0.45 for the AD samples. The first part of the kinetics, as it occurs rapidly, has not been measured directly so that it has been symbolized by arrows in the figure. It is followed by a long term variation close to a time-logarithm behaviour, with roughly the same slope for all the temperatures. Same behaviour occurs for the H/H+C ~ 0.29 and the D/D+C ~ 0.31 samples, with slightly different slopes. As the Tore Supra samples are inhomogeneous (see Figure 4-a), the points reported in figure 5-c are the mean values of mappings obtained at room temperature, to diminish inhomogeneity effects. A value close to $H_D/H_G = 1$ is reached in 300 minutes for the Tore Supra sample at 600°C whereas it is roughly one hundred times longer at 450°C and, after ~21 days, $H_D/H_G$ reaches only 0.7 at 300°C. Many different laws can be used to fit more or less conveniently the data. Then, to really obtain physical information from that data, a complex model, taking into account fundamental phenomena (at least rate equations, at best molecular dynamics) is needed and is not under the scope of this work. For our purpose we decided to fit only the part of the kinetic which can be approximated by a straight line in figures 5-a, b and c (semi-logarithmic scale). The law we chose is the simplest one, not necessarily driven by a physical reason: $H_D/H_G = H_D/H_G|_0 + b \times \log_{10} t/t_0$, where $t_0$ has been taken to 1 minute.

Figure 6 displays the fitting parameter $H_D/H_G|_0$ of the previous empirical relation obtained on the H/H+C ~ 0.29, D/D+C ~ 0.31 and 0.33 samples and the Tore Supra samples as a function of the temperature. $H_D/H_G|_0$, i.e. the value of $H_D/H_G$ at $t=t_0$, probes the rapid initial phase. It is found to increase when the temperature increases for all the three reference samples. This is also true for the Tore Supra sample, even if in the same thermal range the effect is less pronounced, may be due to a higher AD $H_D/H_G$ value for this sample than for that of the reference samples. The less thermally stable sample here is that with initialD/D+C ~ 0.33, as it evolves more rapidly than the others. The data points corresponding to the linear ramp heating have been displayed on the same figure for comparison. There is a quite good superposition between the curves obtained with the linear ramp and the isotherms. It then can lead to the conclusion that the mechanisms responsible for the rapid variation occurring at the beginning of the isotherms, whatever complex they are, are also responsible for the material evolution under linear ramp heating. The need for a kinetic model taking into account the thermally activated processes that occur(clustering of $sp^2$ carbon, scission of $sp^3$ carbon-hydrogen bonds and formation of $sp^2$ carbon, direct transformation of $sp^3$- to $sp^2$-hybridized carbon), like the one presented in [37], is thus needed to access to more detailed information.

## 4. Conclusion

This work reports on the study of the thermal stability of thin films compared to the thermal stability of carbon deposits collected in the Tore Supra tokamak, using linear ramp heating and long term isotherms (from several minutes to 21 days), by means of *in-situ* Raman microscopy. We analyze the structure variation induced by thermal heat treatment (room temperature-800°C) of~200 nm a-C, a-C:H and a-C:D layers with hydrogen and deuterium contents ranging from 29% to 38 % (from hard to soft thin layers) by means of Raman spectroscopy. We confirmed that the layers with the highest H or D content exhibited the poorest thermal stability. We found the same mechanisms to be responsible for the rapid variation found at the beginning of the isotherms, as evidenced in a previous work, and for the material evolution under linear ramp heating. The thermal stability of the Tore Supra deposits shows common trends with those of the reference samples, behaving more like soft layers whereas their initial deuterium content is low. This work can be useful to obtain the evolution of H content or of the complex part of the refractive index using Raman spectra, revealing many correlations between the spectroscopic parameters. For example, from the temperature sensitivity of the G band wavenumber, we extracted information on the $sp^2$ clustering by comparing its thermal behavior to the photon absorption behavior of these layers, at the laser wavelength (514.5 nm). We showed that for the same hydrogen content, the deuterated amorphous carbon layers are less thermally stable than the hydrogenated ones, due probably to growth defects appeared during layer synthesis, associated with the higher momentum of the heavier deuterium.

We have shown here that our methodology leads to an efficient non-contact means for hydrogen isotope release study, which could be helpful in future for analyzing the hydrogenated beryllium deposits in the tokamak ITER.

**Acknowledgments**

C. P. wants to thank W. Jacob, T. Schwarz-Selinger and C. Hopf for providing samples and H or D content analyses.

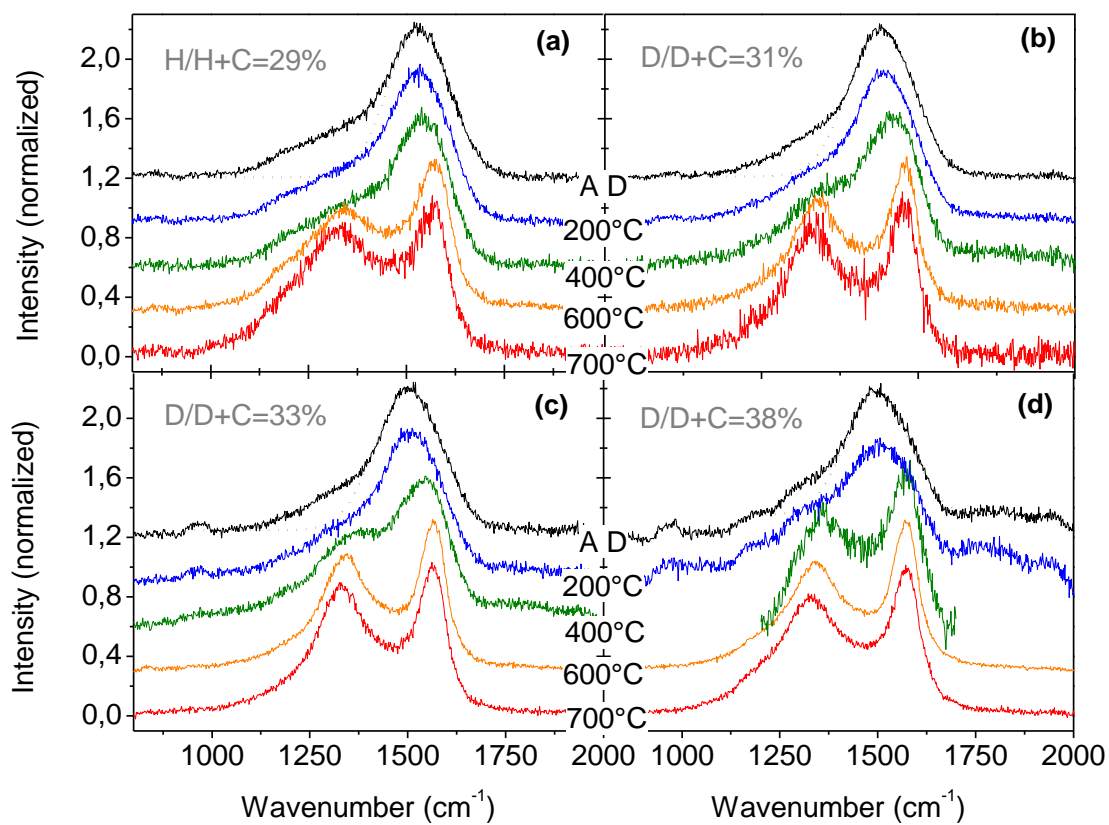

Figure 1. Raman spectra of the 200, 400, 600 and 700°C heated samples recorded during a 3°C.min$^{-1}$ linear Ramp, and compared to their respective as deposited samples (AD). (a) H/H+C=29%. (b) D/D+C=31%. (c) D/D+C=33%. (d) D/D+C=38%. A linear background has been subtracted, spectra were normalized to the height of the G band and then shifted vertically.

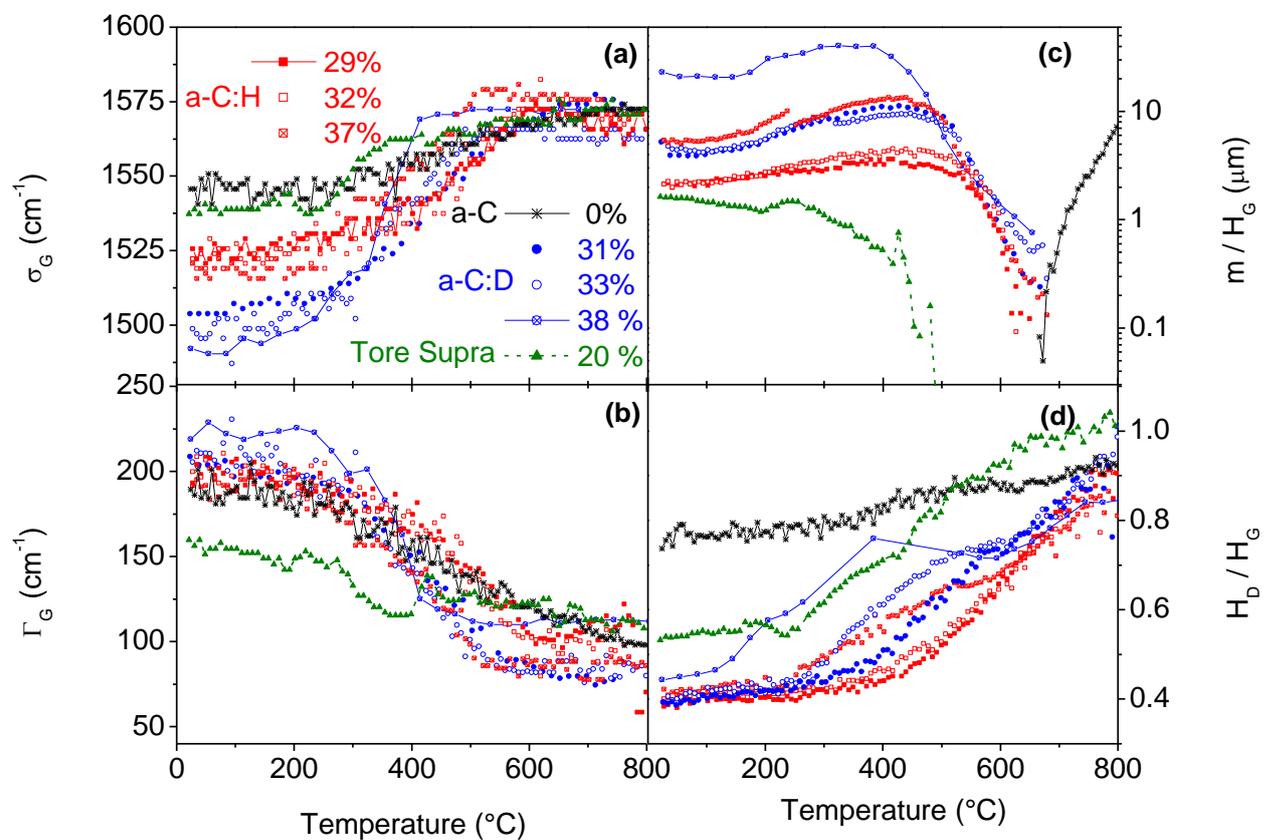

Figure 2. Temperature evolution of Raman parameters of the H+H/C=0.29, 0.32 and 0.37 a-C:H, D+D/C=0.31, 0.33 and 0.38 a-C:D and H-free a-C. (a) G band position $\sigma_G$, (b) G band width $\Gamma_G$, (c) $m/H_G$ parameter, m being the slope of the photoluminescence background and (d) D and G band height ratio $H_D/H_G$. Heating slope is 15 K.min$^{-1}$.

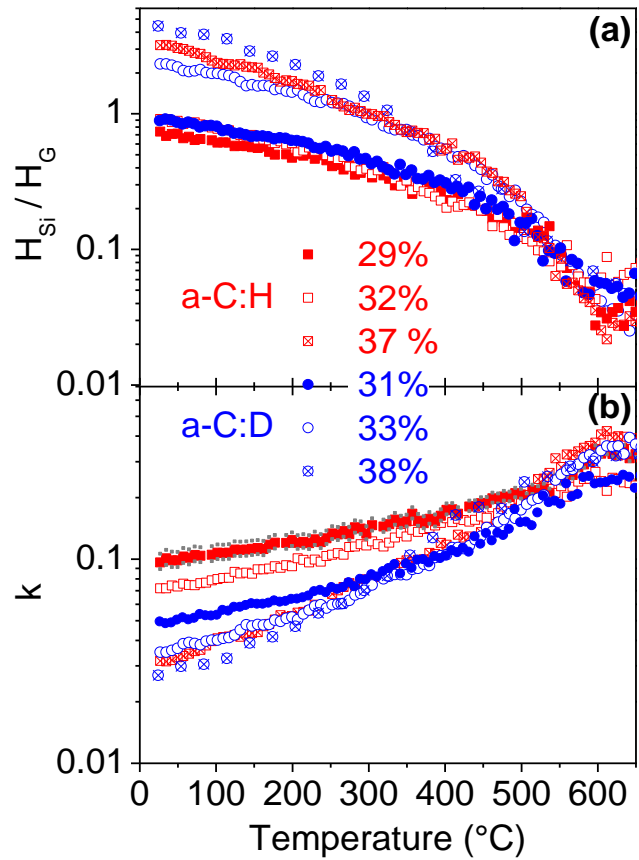

Figure 3. Light absorption properties at 514 nm for the a-C:H and a-C:D layers. (a) Ratio of the silicon over G band height. (b) Complex part of the refractive index k deduced from $H_{Si}/H_G$ according to equation 4 of [36].

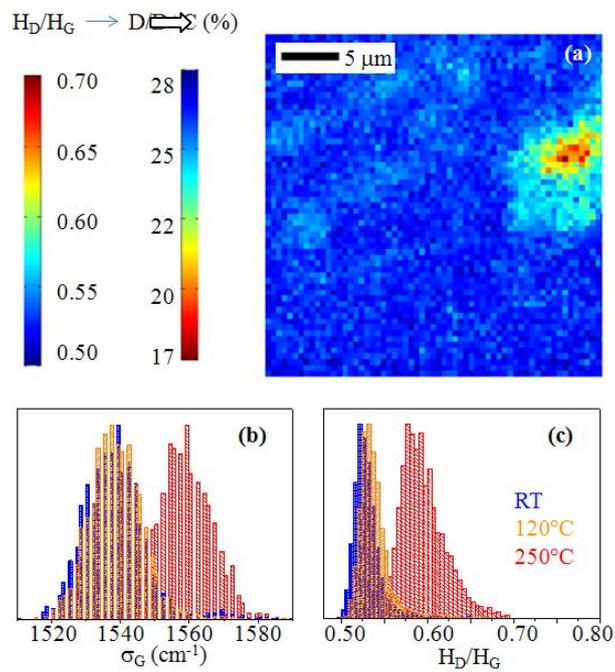

Figure 4. Raman spectroscopy imaging of Tore Supra (TS) samples. (a) $H_D/H_G$ mapping of a unheated TS sample. (b) $\sigma_G$ histograms of unheated and heated TS samples (120, 250°C). (c) $H_D/H_G$ histograms of unheated and heated TS samples.

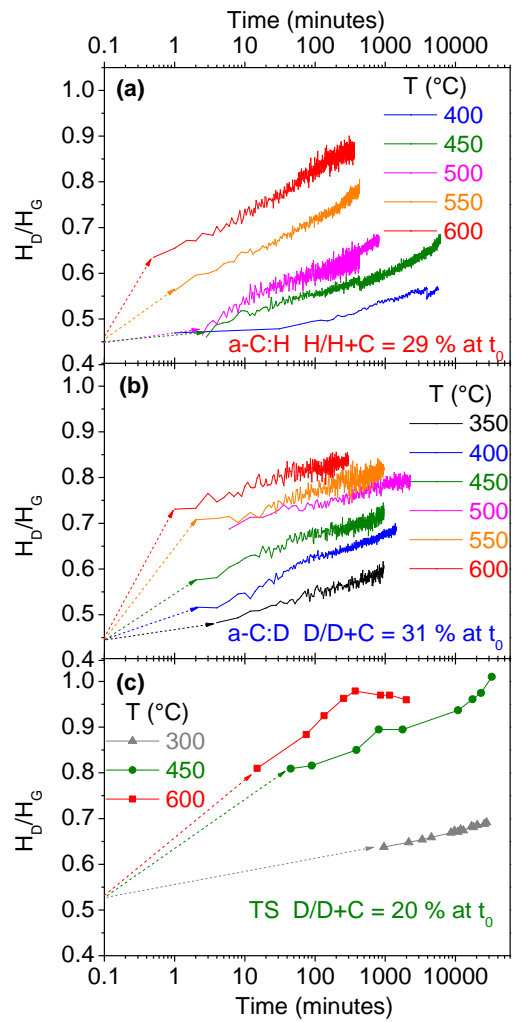

Figure 5. $H_D/H_G$ isotherms. (a) a-C:H sample with 29% of initial H content. (b) a-C:D sample with 31% of initial D content. (C) Tore Supra samples with roughly 20 % of initial D content. (a) and (b) measurements are performed in-situ whereas (c) measurements are performed at room temperature.

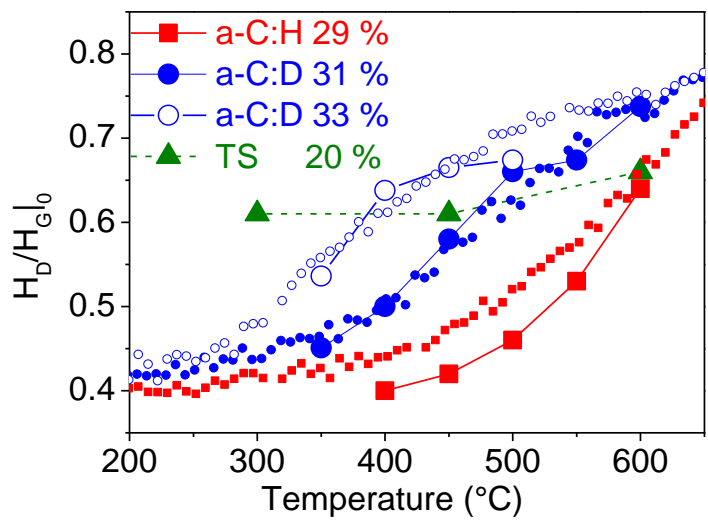

Figure 6. Value at t=t0 of isotherms for the spectral parameter $H_D/H_G$. For comparison, the points during linear ramps (see figure 2-d) are also plotted with a smaller size.

| Sample | DC bias (V) | k (633 nm) | e (nm) | H or D content (at. %) |
|---|---|---|---|---|
| a-C:H | -300 | 0.097 | 240 | 29 |
| a-C:H | -200 | 0.072 | 250 | 32 |
| a-C:H | -100 | 0.032 | 190 | 37 |
| a-C:D | -300 | 0.05 | 210 | 31 |
| a-C:D | -200 | 0.035 | 190 | 33 |
| a-C:D | -100 | 0.027 | 210 | 38 |

Table 1. a-C:H and a-C:D layer properties. e is the thickness and k the complex part of the refractive index determined at 633 nm. See the experimental section for details.